**OPEN**

# Critical Transitions in Intensive Care Units: A Sepsis Case Study

Pejman F. Ghalati[1], Satya S. Samal[1,3], Jayesh S. Bhat[1], Robert Deisz[2], Gernot Marx[2] & Andreas Schuppert[1]

The progression of complex human diseases is associated with critical transitions across dynamical regimes. These transitions often spawn early-warning signals and provide insights into the underlying disease-driving mechanisms. In this paper, we propose a computational method based on surprise loss (SL) to discover data-driven indicators of such transitions in a multivariate time series dataset of septic shock and non-sepsis patient cohorts (MIMIC-III database). The core idea of SL is to train a mathematical model on time series in an unsupervised fashion and to quantify the deterioration of the model's forecast (out-of-sample) performance relative to its past (in-sample) performance. Considering the highest value of the moving average of SL as a critical transition, our retrospective analysis revealed that critical transitions occurred at a median of over 35 hours before the onset of septic shock, which suggests the applicability of our method as an early-warning indicator. Furthermore, we show that clinical variables at critical-transition regions are significantly different between septic shock and non-sepsis cohorts. Therefore, our paper contributes a critical-transition-based data-sampling strategy that can be utilized for further analysis, such as patient classification. Moreover, our method outperformed other indicators of critical transition in complex systems, such as temporal autocorrelation and variance.

Certain biological systems exhibit nonlinear dynamics that undergo sudden regime transitions at tipping points[1,2]. In a medical context, these transitions often indicate changes in clinical phenotypes, e.g., disease-onset[3]. Such phenomena have been studied mathematically with techniques from the application of singularity theory to dynamical systems[4–6]. In addition, data-driven methods use statistical indicators known as early-warning signals to model the dynamics of systems approaching transitions[7–14]. Modeling such transitions is beneficial for several applications in systems medicine, such as monitoring health[15,16], predicting disease-onset and gaining an improved understanding of the underlying disease progression[17].

Our focus is on sepsis, a common complication in the intensive care unit (ICU), and we introduce a notion of regime transition in septic dynamics. As stated in the Third International Consensus Definitions of Sepsis and Septic Shock (Sepsis-3), "sepsis is a life-threatening organ dysfunction caused by a dysregulated host response to infection", and "septic shock is a subset of sepsis in which underlying circulatory and cellular/metabolic abnormalities are profound enough to substantially increase mortality[18]". Sepsis causes a high rate of in-hospital mortality and costs the healthcare sector billions due to rising incidence rates and prolonged hospital stays[19,20]. Accurate diagnosis, however, remains a challenging task for physicians due to the heterogeneity of infectious agents and the frequent presence of multiple comorbidities. Early, aggressive administration of antibiotics is crucial, and delays in this treatment significantly increase mortality[21,22].

To detect signs of sepsis early, numerous illness severity scores or early-warning signals exist: the Acute Physiology and Chronic Health Evaluation (APACHE II), the Simplified Acute Physiology Score (SAPS II), the Sepsis-related Organ Failure Assessment Score (SOFA), the Modified Early Warning Score (MEWS), and the Simple Clinical Score[23]. These scores are good predictors of general disease severity and mortality but cannot estimate the risk of developing sepsis with reasonable sensitivity and specificity[23].

Numerous machine learning (ML) methods were therefore developed to predict sepsis onset[24–26]. Rothman et al.[27] used structured information from electronic health records (EHRs) to identify sepsis on admission or to predict its onset during hospitalization. For septic shock prediction, Ghosh et al.[28] proposed an integrative model combining sequential contrast patterns with coupled hidden Markov models. Henry et al.[23] developed a targeted real-time early-warning score (TREWScore) by training a Cox regression model to identify patients at high risk of developing septic shock. Additionally, Horng et al.[29] argued that combining free-text patient data with other predictor features significantly improved the performance of ML models. Although these ML approaches have the potential to increase diagnostic accuracy, they involve time-consuming and domain-specific variable/feature selection[30,31]. Our proposed method can be considered in the preprocessing stages to select appropriate data for further downstream analysis.

[1]Joint Research Center for Computational Biomedicine, RWTH Aachen University, 52074, Aachen, Germany. [2]Klinik für Operative Intensivmedizin and Intermediate Care, Universitätsklinikum Aachen, 52074, Aachen, Germany. [3]Present address: BASF SE, Carl-Bosch-Strasse 38, 67056, Ludwigshafen am Rhein, Germany. Pejman F. Ghalati and Satya S. Samal contributed equally. Correspondence and requests for materials should be addressed to A.S. (email: schuppert@combine.rwth-aachen.de)

Our computational method aims to identify and characterize signals indicative of critical transitions based on the concept of surprise loss (SL)[32]. SL was originally developed in econometrics to assess forecast breakdown, i.e., instability in the model's forecasting ability. Such instability was attributed to instability in the underlying data-generating process, whose effects have been studied from a mathematical perspective[33,34]. We assume that similar instability occurs in patient data because of changes in the underlying biological mechanism due to medical intervention or disease progression.

We utilize SL to identify regions in the time series where the data-generating process changes and quantify them with a numerical score. The score captures the extent of deviation between the past performance of a model and its future performance. We consider the highest value of such a score to be a putative tipping point in the disease dynamics, and we consider it as a surrogate for critical transition. In addition, we present a critical-transition-based data-sampling strategy is also presented where data are sampled at regions around critical transition; this strategy outperforms random sampling in differentiation between septic shock and non-sepsis patients. We also compare our approach to methods based on autocorrelation and variance[7,15,16,35], which have been used to identify early-warning signals of critical transitions.

## Materials and Methods

**Data source.** We sourced patients' multivariate time series data from the publicly available EHR database, Medical Information Mart for Intensive Care MIMIC-III v1.4[36], which contained longitudinal data of 46,520 deidentified patients from 58,976 distinct ICU admissions. For ease of interpretation, we treated each admission as a distinct patient.

In the ICU, clinical staff make swift decisions or take prompt actions during patient management. These employees prioritize timely and correct treatment over consistent documentation of their processes, thereby limiting the reliability of clinical annotation for retrospective analysis. Furthermore, the execution of guidelines for identifying imminent disease varies across hospitals. Hence, we restricted our data analysis to predominantly machine-recorded quantitative variables.

Decision rules for retrospective annotation of the sepsis syndrome have evolved over the decades as knowledge of its pathophysiology and epidemiological impact have increased[37]. Whereas earlier definitions (1991[38], 2001[39]) focused on uncontrolled systemic inflammation as the major indicator, the latest 2016[18] definition, commonly known as Sepsis-3, emphasizes organ dysfunction as the leading effect of the sepsis syndrome and proposes to update the International Classification of Diseases (ICD) coding system[40,41] (ICD-9: 995.92, 785.52; ICD-10: R65.20, R65.21). SOFA scoring system grades the extent of organ dysfunction and is calculated every 24 hours during a patient's ICU stay[42,43].

Because the ICD-9 codes in our data were not compatible with Sepsis-3, we annotated the patient data in accordance with Table 2 from the 2016 consensus definition[18]. Fig. 1 illustrates a general schematic of our annotation framework.

The annotation framework was applied to all 58,976 patients, identifying 22,547 (38.2%) sepsis patients and 3208 (5.4%) septic shock patients. Among the 3208 septic shock patients, we analyzed only adults (18+ years old at admission) with at least a 36-hours stay and at most 144 hours spent in the unit before onset, which generated a cohort of 630 patients. Our non-sepsis cohort comprised 6,236 patients who lacked Sepsis-3 annotation or sepsis-specific ICD-9 codes and who stayed between 36 and 144 hours in the ICU. Demographic information on the two cohorts can be found in Supplementary Table S1.

We cannot exhaustively analyze and validate the accuracy of our annotation framework owing to the absence of a manually curated "ground truth" dataset of Sepsis-3 patients. Software implementations with different data cleaning processes and patient exclusion criteria (PEC) from the same annotation framework could result in divergent cohorts. For example, for the same database, another implementation[44] annotated almost half (49.1%) of their analysis cohort (n = 11,791; reasonable PEC) as Sepsis-3, whereas our implementation annotated approximately 38% of the entire population (n = 58,976; no PEC). There may be a high degree of overlap in the annotated cohorts; thus, a comparison of the two implementations is currently under way.

Based on availability and relevance to sepsis, we preselected groups of variables: the laboratory variables included bicarbonate, creatinine, blood urea nitrogen (BUN), hematocrit, hemoglobin, platelet count, white blood cell count (WBC), potassium, and sodium; the vital signs and physiological variables comprised body temperature, heart rate, respiratory rate, oxygen saturation (SpO2), arterial blood pressure (systolic, mean, and diastolic), and urine output; the two septic markers comprised the shock index (ratio of heart rate over systolic blood pressure), and the ratio of BUN to creatinine[23]. Table 1 shows the mean sampling rates of the variables in the respective patient cohorts, and their distribution can be seen in Supplementary Fig. S1.

*Missing value imputation and time binning.* Data representation is a crucial step in analyzing time series. Continuous EHRs suffer from missing values due to insufficient data collection and lack of documentation. Additionally, high heterogeneity in variable type and irregular sampling intervals make such data difficult to handle. To address the problems of missing data and data sparsity, we transformed our time series into 30-minute time bins by imputing values in the bins and averaging measurement values over the bins. We experimented with different imputation methods, such as linear, polynomial and Stineman interpolation[45]. The Stineman method was chosen due to its superior performance in reducing overshoots and handling sharp changes in the imputed values.

| Variable | Sampling Rate per Day | |
| | Septic Shock | Non-Sepsis |
|---|---|---|
| BUN | 1.7 | 1.1 |
| Creatinine | 1.7 | 2.6 |
| Hemoglobin | 1.9 | 5.0 |
| Bicarbonate | 1.7 | 1.5 |
| Respiratory Rate | 20.5 | 4.4 |
| Heart Rate | 20.2 | 9.8 |
| Hematocrit | 2.2 | 8.2 |
| WBC | 1.5 | 9.2 |
| SpO2 | 20.1 | 8.1 |
| Platelets | 1.6 | 5.3 |
| Systolic BP | 19.8 | 9.3 |
| Urine Output | 12.1 | 10.2 |
| Temperature | 6.7 | 9.1 |
| Sodium | 2.0 | 3.2 |
| Diastolic BP | 19.8 | 10.4 |
| Mean BP | 20.0 | 6.6 |
| Potassium | 2.5 | 3.8 |

**Table 1.** Mean sampling rate of the preselected clinical variables in the septic shock and non-sepsis cohorts. For a single patient, the sampling rate was the ratio of the number of observations recorded in the ICU to the length of the patient's stay (in days). The rate was then averaged over all the patients in the respective cohorts.

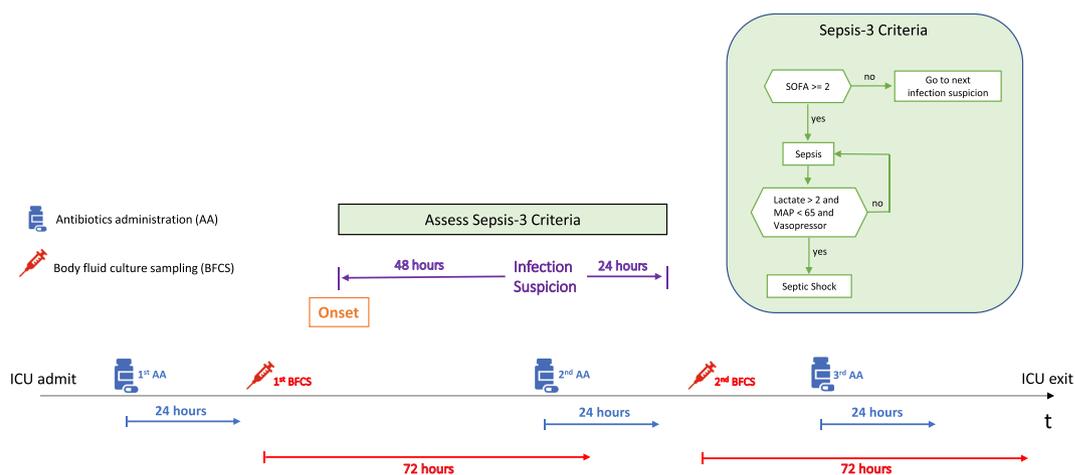

**Figure 1.** Over the length of a patient's ICU stay, all timestamps of body fluid (blood, urine, cerebrospinal fluid) sampling and antibiotic administration were retrieved. For each of the timestamps, an infection was suspected if antibiotics were administered within 72 hours of any prior body fluid sampling (irrespective of culture findings) or if any body fluids were sampled within 24 hours of prior antibiotic administration. Sepsis-3 criteria were independently evaluated over time windows around the infection-suspected timestamps (IST). Each time window began 48 hours prior to IST until 24 hours post IST. If the criteria were satisfied during a given time window, then the beginning of the window was annotated as the onset time. In the schematic, the 2nd antibiotics administration falls within 72 hours of previous body fluid sampling; thus, an infection is suspected.

*Data normalization.* Our variables (Table 1) had different scales and measurement units. Data normalization was therefore needed for our method. For this purpose, we transformed the observables by Z-score normalization to address the use of different units of measurement.

**State space model.** To define SL, we require a dynamical mathematical model for our multivariate clinical time series. Here, we consider a state space model (SSM) approach[46,47], which models the data in a hierarchical manner with hidden states that give rise to observables. In our context, the hidden states can be assumed to represent the biological processes, and the observables represent the clinically measured variables. The observables in our SSM are expressed as linear combinations of hidden random states. Such a model incorporates the variations

| Variable | BF (SL) | BF (AC1) | BF (VAR) |
|---|---|---|---|
| BUN | **0.98** | 0.88 | 0.95 |
| BUN-Creatinine | **1.00** | 0.20 | 0.43 |
| Creatinine | 0.94 | 0.98 | **1.00** |
| Hemoglobin | **0.86** | 0.85 | 0.07 |
| Hematocrit | **0.99** | 0.81 | 0.09 |
| Shock Index | **0.99** | 0.03 | 0.72 |
| Respiratory Rate | 0.75 | **0.82** | 0.61 |
| Heart Rate | **0.75** | 0.09 | 0.62 |
| Systolic BP | **1.00** | 0.03 | 0.40 |
| Bicarbonate | 0.1 | 0.89 | **1.00** |
| Platelets | **0.35** | 0.15 | 0.08 |
| Temperature | **0.99** | 0.86 | 0.93 |
| Urine Output | 0.20 | **0.97** | 0.66 |
| WBC | 0.75 | 0.97 | **1.00** |
| Mean BP | **0.97** | 0.05 | 0.37 |
| SpO2 | **0.91** | 0.89 | 0.71 |
| Diastolic BP | **0.85** | 0.06 | 0.57 |
| Sodium | **0.42** | 0.34 | 0.16 |
| Potassium | **0.82** | 0.04 | 0.13 |

**Table 2.** A statistical significance test (see 'Data-sampling strategy with SLMean') was performed to test whether the values of clinical variables at largest *SLMean*, *AC1* and *VAR* were able to statistically differentiate septic shock patients from non-sepsis patients, and a bootstrap test was performed to calculate the fraction of replications where the $t_{max}$ p-values were less than the p-values from a random sampling. The columns of the table show the result achieved by each variable: BF (SL), BF (AC1) and BF (VAR). The highest BF for each clinical variable is shown in bold.

- - -

in the biological processes and a measurement noise term. The variations due to biology are modeled by adding a stochastic term to the hidden states, whereas the measurement noise term is added to the observables. Both terms are assumed to follow a multivariate normal (MVN) distribution.

The computation of SL is agnostic to the underlying dynamical model. The SL literature[32] uses a linear dynamical model, whereas we use an SSM for our application. The primary reason to use this type of model is to separate the biological processes from the observables, i.e., to model two sources of variability. Below, we represent such an SSM model.

$$
\begin{aligned}
x_t &= x_{t-1} + w_t & \text{where } w_t \sim MVN(0, Q), \ x_0 \sim MVN(\pi, \wedge) \\
y_t &= Zx_t + a + v_t & \text{where } v_t \sim MVN(0, R)
\end{aligned}
\tag{1}
$$

where the indices of the time series are from $t = 1, \ldots, T$; $e$ is the number of hidden trends; $x$ is an $e \times T$ matrix of hidden states; $y$ is an $n \times T$ matrix of $n$ observables; and $w$ is an $e \times T$ matrix of process error. In general, $e \ll n$. The process error at time $t$ follows an MVN distribution with mean 0 and $e \times e$ covariance matrix $Q$; $v$ is an $n \times T$ matrix of observation error. The observation error at time $t$ follows an MVN distribution with mean 0 and $n \times n$ covariance matrix $R$; $Z$ is an $n \times e$ parameter matrix; $a$ is a vector of offsets; $\pi$ is a matrix of $e \times 1$ means; $\wedge$ is an $e \times e$ covariance matrix. The set of parameters can be represented in compact form as $\theta = (Q, R, Z, x_1, \ldots, T, \pi, \wedge)$, and their estimate is $\hat{\theta}$. $\hat{y}_t$ and $\tilde{y}_{t+\lambda}$ are the estimate and $\lambda$-step-ahead forecast, respectively, of the given observables $y_t$.

Our implementation incorporated MARSS[48,49], which is an R package for fitting constrained and unconstrained linear multivariate autoregressive SSMs by maximum likelihood parameter estimation. We utilized MARSS to fit an SSM to our multivariate time series data, using its recommended initial conditions that ensure parameter identifiability. We assumed the presence of multiple hidden states and fixed $e = 3$. Furthermore, we evaluated the robustness of our results with respect to the changes in the model parameters (see 'Robustness of the SSM model').

**Perturbations in the dynamics.** *Early-warning indicator.* Our proposed computational method based on surprise loss (SL)[32] computes the difference between the forecast error, i.e., out-of-sample error, and the in-sample performance. The out-of-sample error measures the quality of model forecasts, i.e., the prediction of the model for the data that were not used for fitting, whereas the in-sample error quantifies the deviation between the model estimates and the data that were used for model fitting. A high out-of-sample error compared to the in-sample error is suggestive of instability in the patient data. In such a scheme, our model may be a poor fit for the data, but we are interested in evaluating whether the past performance of the model is consistent with future forecasts. The performance is measured for a fixed loss function using a moving time window. Furthermore, the SL computation is unsupervised, i.e., the clinical conditions of patients, such as septic shock or non-sepsis, are not required. Originally, the idea of SL was used to perform a statistical test to determine forecast breakdown in time series, i.e., to determine whether the average of SL is close to zero[32]. However, in our application, the aim is not to test whether a given time series underwent a statistically significant forecast breakdown; rather, it is to identify high SL values in the given time series and later use this information in postprocessing steps (see 'Data-sampling strategy with SLMean').

In spirit, this approach is close to the identification of structural breaks or change-points analysis[50,51]. However, the SL-based approach has the additional advantage of being robust to model misspecification. Specifically, in practice, the SSM model (i.e., the functional form and variables) is likely to be misspecified and may not be a good approximation of the underlying disease processes. By formalizing SL as the difference between in-sample and out-of-sample performance and not relying on model parameters or error variances, the SL-based approach provides a natural way to handle such scenarios (see 'Relationship with the literature' in Giacomini et al.[32]).

With a moving time window of width $m$, the SSM model (see equation (1)) was fitted for time indices $t - m + 1, \ldots, t$. $y_t^{i_c}$ denotes the observables of a given patient $i$ with clinical condition $c$ at time index $t$, and $T^{i_c}$ is the length of the corresponding time series. The in-sample error is a quadratic loss function that averages the squared differences between the estimated and the given observables, and it is denoted as $L_j(\hat{\theta}_t^{i_c}) = \frac{1}{n}\sum_{k=1}^{n}(y(k)_j^{i_c} - \hat{y}(k)_j^{i_c})^2$ where $y(k)_j^{i_c}$ is the $k^{th}$ element of column vector $y_j^{i_c}$. Similarly, the out-of-sample error is a quadratic loss function that averages the squared differences between the $\lambda$-step-ahead forecast and the given observables, and it is denoted as $L_{t+\lambda}(\hat{\theta}_t^{i_c}) = \frac{1}{n}\sum_{k=1}^{n}(y(k)_{t+\lambda}^{i_c} - \tilde{y}(k)_{t+\lambda}^{i_c})^2$. The SL is the difference between the out-of-sample and the in-sample error:

$$SL_{t+\lambda}^{i_c} = L_{t+\lambda}(\hat{\theta}_t^{i_c}) - \frac{1}{m}\sum_{j=t-m+1}^{t} L_j(\hat{\theta}_t^{i_c}) \qquad \text{for} \quad t = m, \ldots, T^{i_c} - \lambda \tag{2}$$

To remove short-term fluctuations, a moving-average filter (with size $\delta$) smooths the SL:

$$SLMean_t^{i_c} = \sum_{j=t-(m+\delta)+1}^{t} \frac{SL_j^{i_c}}{j} \qquad \text{for} \quad t = m + \delta, \ldots, T^{i_c} \tag{3}$$

For a given patient $i$, prior to the clinically annotated onset of disease $c$, a relatively high $SLMean^{i_c}$ suggests putative transitions across dynamical regimes and serves as an early-warning indicator. We consider the maximum of $SLMean^{i_c}$ at time index $t_{max}^{i_c}$ to denote a critical transition. Fig. 2a illustrates the calculation of $SL^{i_c}$, $SLMean^{i_c}$ and $t_{max}^{i_c}$. A simulated example using synthetic data is shown in Fig. 3.

*Uncertainty in SLMean.* Uncertainty in out-of-sample forecasting and in-sample performance adds noise to the precise location of $t_{max}^{i_c}$. Let $t_{max(up)}^{i_c}$ and $t_{max(low)}^{i_c}$, respectively, be the time indices corresponding to the modes of the upper and lower bounds of the 95% prediction interval of $SLMean$. Our approach is robust if the deviations of $t_{max}^{i_c}$ from $t_{max(up)}^{i_c}$ and $t_{max(low)}^{i_c}$ are close to zero.

### Data-sampling strategy with *SLMean*.
Here, we demonstrate a method for sampling data from the critical transition points (derived from *SLMean*) to differentiate the septic shock cohort from the non-sepsis cohort (see Fig. 2b). We also propose a bootstrap test (based on a random sampling of data) to evaluate whether it outperforms the SL-based approach. Such a data selection step can be seen as a preprocessing step for the machine learning-based techniques being developed to study sepsis (as described in 'Introduction'). The data sampling step is agnostic to the clinical condition of the patient, i.e., data for each patient are based on SL (see 'Perturbations in the dynamics'), and in a subsequent step, we used the clinical condition to perform statistical tests.

Specifically, we selected the data at $t_{max}^{i_c}$, i.e., the critical transition points (in the case of multiple $t_{max}^{i_c}$ values, the one closer to the disease-onset was selected), sampled the corresponding data and represented them as an $n \times v$ variable matrix $S^c = [y_{t_{max}^1}^{1_c}, \ldots, y_{t_{max}^v}^{v_c}]$ where $c \in \{0, 1\}$ i.e., non-sepsis and septic shock conditions, and $v$ is the total number of patients. Thereafter, for each variable, a p-value based on Wilcoxon rank-sum test[52] was calculated, quantifying the significance of differences between the two patient cohorts (as shown in the equation (4)).

$$p = (pval(S_1^0, S_1^1), \ldots, pval(S_n^0, S_n^1)) \tag{4}$$

where $pval(.)$ returns the p-value based on the Wilcoxon rank-sum test. $S_j^0$ and $S_j^1$ denote the $j^{th}$ row vectors of matrices $S^0$ and $S^1$ matrices, respectively. Furthermore, we performed the Benjamini and Hochberg correction method to adjust the p-values[53] accounting for multiple comparisons.

### Bootstrapping.
Furthermore, a bootstrap test was used to compare the p-values calculated at critical transition points from the p-values that were obtained from random points in our time series. For a randomly selected time index $t$ with its corresponding observation $y_t^{i_c}$, where $t \in (1, T^{i_c})$, the $t_{random}$ p-values were calculated by replacing $t_{max}$ with $t$. The test was repeated 1000 times. Bootstrap frequency (BF) denotes the fraction of replications wherein $t_{max}$ p-values were less than $t_{random}$ p-values. A high BF value indicates that the SL based approach has an advantage over the random approach. In addition to computing BF on data randomly sampled from all times, we computed BF on randomly sampled data of septic patients from two arbitrary time intervals, 36 hours and 18 hours before the onset of septic shock. This step allows us to test whether merely randomly sampling data close to the onset time can outperform the SL approach.

### Autocorrelation and variance as early-warning signals.
In the dynamics of a system, increased temporal autocorrelation and increased variance are hypothesized to be two indicators that the system is approaching a state transition[7]. To evaluate the SL concept, we calculated these two presumed early warning signals and compared the results with those obtained from the SL approach. As these measures are both univariate, to apply them to our multivariate time series data, we formulated them as follows:

$$AC1_t^{i_c} = \frac{1}{N} \sum_{k=1}^{N} AC(y(k)_t)^{i_c} \qquad \text{for} \quad t = m, \ldots, T^{i_c} \tag{5}$$

$$VAR_t^{i_c} = \frac{1}{N} \sum_{k=1}^{N} Var(y(k)_t)^{i_c} \qquad \text{for} \quad t = m, \ldots, T^{i_c} \tag{6}$$

where $AC$ and $AC1$ are autocorrelation and variance functions applied on variable $y(k)$ for time indices $t - m + 1, \ldots, t$. $t$ is the time index, and $m$ is the width of a moving time window. The first coefficient of auto-correlation $AC1_t^{i_c}$ and variance $VAR_t^{i_c}$ were computed by averaging over $N$ variables. $i$ is the index of a given patient with clinical condition $c$, and $T^{i_c}$ is the length of the corresponding time series.

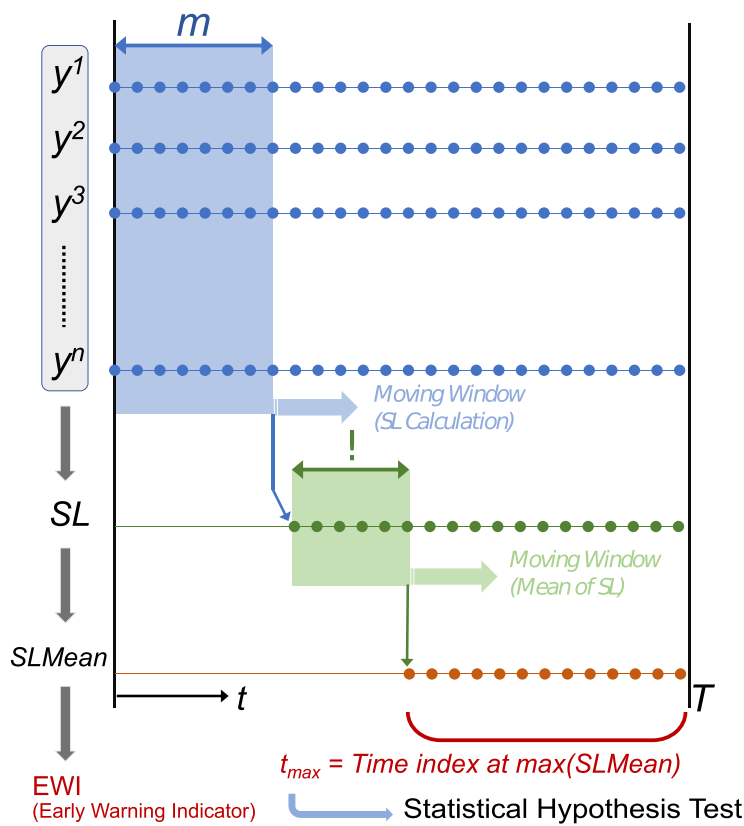

(a)

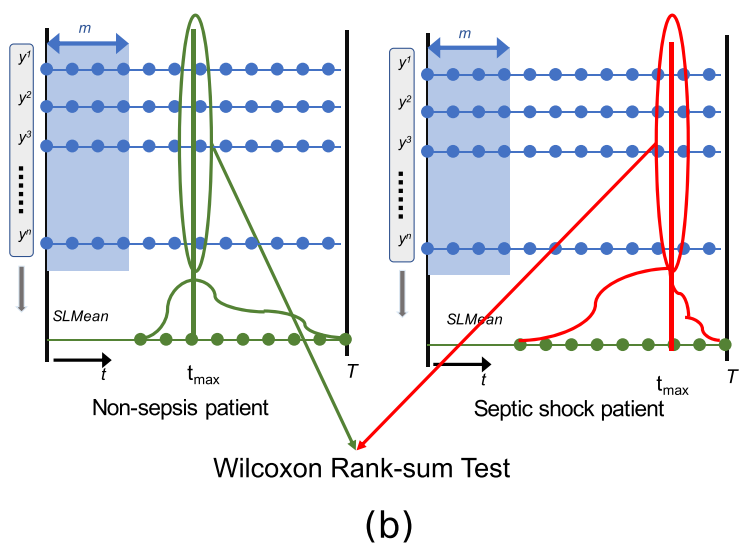

(b)

**Figure 2.** (**a**) A schematic for the calculation of $SL^{i_c}$, $SLMean^{i_c}$, and $t_{max}^{i_c}$ for a given patient $i$ and clinical condition $c$. The SSM was fitted with a moving time window of length $m$ (as shown in blue) and the $SL^{i_c}$ was calculated. A second sliding window of length $\delta$ was used to compute the $SLMean^{i_c}$ (as illustrated in green). The $T^{i_c}$ denotes disease onset in septic shock patients and it represents the time of discharge or death in non-sepsis patients. The $t_{max}^{i_c}$ denotes the time index of the highest $SLMean^{i_c}$ and it was used in our data-sampling approach. (**b**) A schematic diagram illustrating our data-sampling strategy using our method. Observables at the time of highest $SLMean$ magnitudes, i.e., critical transition points, were selected from septic shock and non-sepsis patients. The Wilcoxon rank-sum test was used to determine the statistical significance of the changes in the observables.

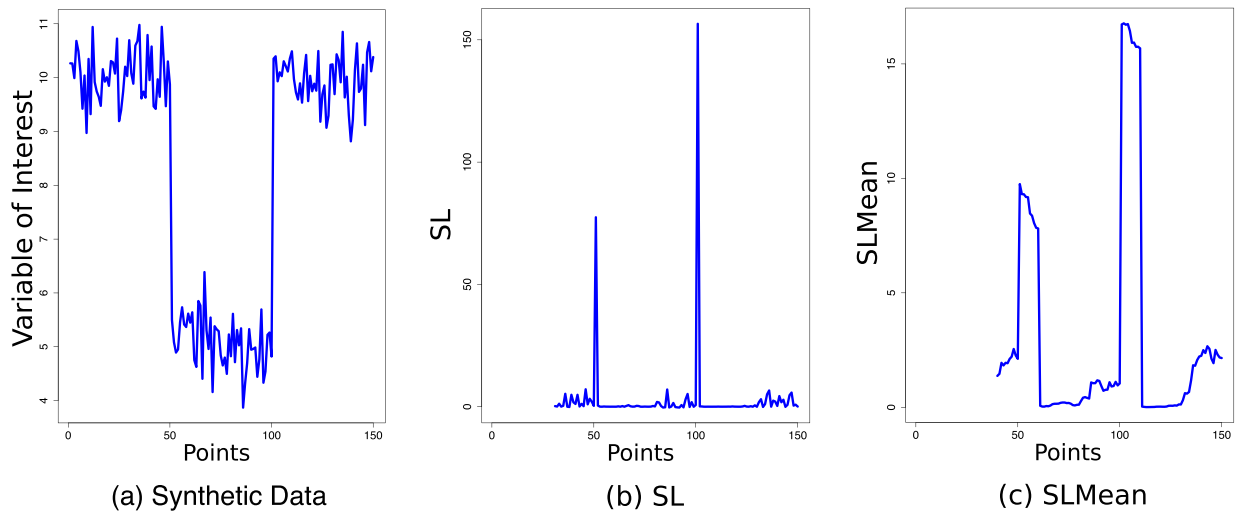

**Figure 3.** Artificial example showing the calculation of *SLMean* from a synthetic dataset that was generated by concatenating 50 points, drawn independently from three univariate normal distributions with different means (5, 10, 15) and a standard deviation of 0.5. Computed with a moving time window of length 30 and the number of hidden states set to 1, the magnitude of SL intensified at the $50^{th}$ and $100^{th}$ time-points, where the parameters of the data-generating process changed, i.e., a proxy for transitions across different dynamical regimes.

Similar to the SL concept, $t_{max}$ is defined as the time index where the highest value of the early-warning signal occurs (here, the largest value of $AC1^{i_c}$ or $VAR^{i_c}$). P-values and bootstrap frequencies were computed as described in 'SLMean-based data-sampling strategy' and 'Data-sampling strategy with SLMean'.

**Software.** To support reproducible research, our computational method is available at https://github.com/JRC-COMBINE/SL-MTS.

## Results

***SLMean* as an early-warning indicator.** Over a moving time window ($m = 36$, i.e., 18 hours; $e = 3$; $\lambda$-step-ahead = 1, i.e., 30 minutes; $\delta = 6$, i.e., 3 hours), the $SLMean^{i_c}$ values ('Perturbations in the dynamics'), as shown in Fig. 4, were computed. A positive $SLMean^{i_c}$ indicates higher out-of-sample error than in-sample error, signaling putative transitions in the underlying dynamics. The componentwise mean vector and associated standard deviation of all septic shock patients, i.e., $SLMean^{1_c}, \ldots, SLMean^{N_c}$ (where $N$ is the total number of septic shock patients and $c$ is the septic shock clinical condition), intensified as the moving time window approached the disease onset. For the same cohort of septic shock patients, a slight increase in the componentwise mean vector and associated standard deviation of $VAR^{i_c}, \ldots, VAR^{N_c}$ could be seen, while those of $AC1^{i_c}, \ldots, AC1^{N_c}$ did not show any changes over time. The findings are summarized in Fig. 5.

It should be taken into account that the largest $SLMean_t^{i_c}$ need not necessarily occur exactly at the time of disease onset. For septic shock patients, the location of the time index $t_{max}$ from the onset time ($T$) is shown in Fig. 6b. In the majority of our patients' data, the highest $SLMean$ occurred near septic shock onset (60% of the patients, the signal occurred less than 48 hours prior to onset, as shown in Fig. 6b). However, in some patients, the signal was observed beyond onset time. Possible explanations include a lack of records or a low sampling rate of variables a few days before the onset of septic shock, resulting in a nonsignificant $SLMean$. The highest $SLMean$, on average, occurred 46 hours (median of 35.6 hours) prior to the appearance of septic shock symptoms. In comparison, TREWScore[23] identified septic patients at a median of 28.2 hours before onset.

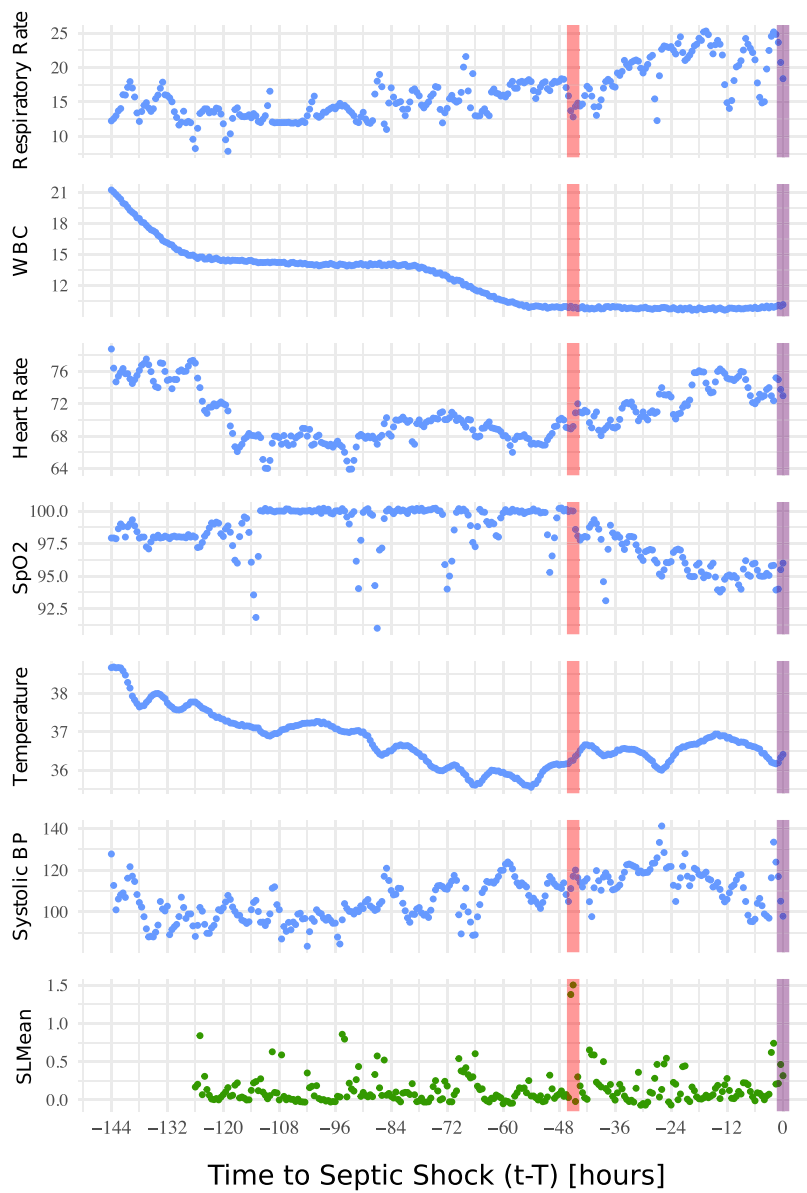

**Figure 4.** The changes over time in a group of clinical variables used in this study and the corresponding computed $SLMean^{i_c}$ of a sample septic patient before the onset of septic shock (violet line). The $SLMean^{i_c}$ is calculated over a moving time window ($m = 36$, i.e., 18 hours; $e = 3$; $\lambda$-step-ahead $= 1$, i.e., 30 minutes; $\delta = 6$, i.e., 3 hours). The red line shows the time location ($t_{max}^{i_c}$) of the largest $SLMean^{i_c}$ (i.e., the critical transition point).

While the median time of the peak $SLMean$ occurred at 35.6 hours before the onset of septic shock, visual inspection of the mean and standard deviation of $SLMean$ indicates an upward trend starting from approximately 24 hours (Figs 5a and 6a). The explanation for the apparent deviation from the baseline is that the highest $SLMean^{i_c}$ values that occurred closer to onset were greater in magnitude.

Furthermore, we determined the uncertainty in SL calculation using prediction intervals (as described in 'Uncertainty in SLMean'). Our results show negligible deviation in $t_{max}$ i.e., the median deviation is 0, and the interquartile range (IQR) is 5.4 hours.

**SLMean-based data-sampling strategy.** We compared the p-values for data sampled at $t_{max}$ (i.e., critical transition point) to those obtained via random sampling (see equation (4) and 'Data-sampling strategy with SLMean'). The same procedure was implemented for $AC1$ and $VAR$, and the bootstrap test was performed for all time indices. The bootstrap frequencies were denoted as $BF$ ($SL$), $BF$ ($AC1$) and $BF$ ($VAR$), respectively (see Table 2). The different $BF$ computations test the association of the bootstrap frequency values of some variables with high $SLMean$, $AC1$ and $VAR$. In 14 out of 19 variables, $BF$ ($SL$) demonstrates superior results. In the next step, in addition to all the time indices, the bootstrap test was performed for time-windows of 18 and 36 hours before the onset of septic shock; the bootstrap frequencies are represented as $BF$ ($Full$), $BF$ ($18\,hours$), and $BF$ ($36\,hours$). Fig. 7a plots $BF$ ($Full$) against p-values computed at $t_{random}$ and at high $SLMean$ (i.e., $t_{max}$). Most of the variables show a good *BF* with high log-transformed p-values when sampled at large *SLMean*, particularly in the case of variables such as blood pressures, temperature and SpO2, where random sampling leads to poor p-values. As the random sampling strategy changed to either to 36 or 18 hours in Fig. 7b, *BF* reduced for six variables (WBC, diastolic blood pressure, Hemoglobin, SpO2, creatinine, and BUN), but it was preserved for nine variables (respiratory rate, heart rate, potassium, mean blood pressure, hematocrit, shock index, temperature, BUN-creatinine, and systolic blood pressure), i.e., the differences among *BF* (*Full*), *BF* (*36 hours*), and *BF* (*18 hours*) were small. Four variables, bicarbonate, urine output, platelets and sodium, had low *BF* (*Full*), *BF* (*36 hours*), and *BF* (*18 hours*).

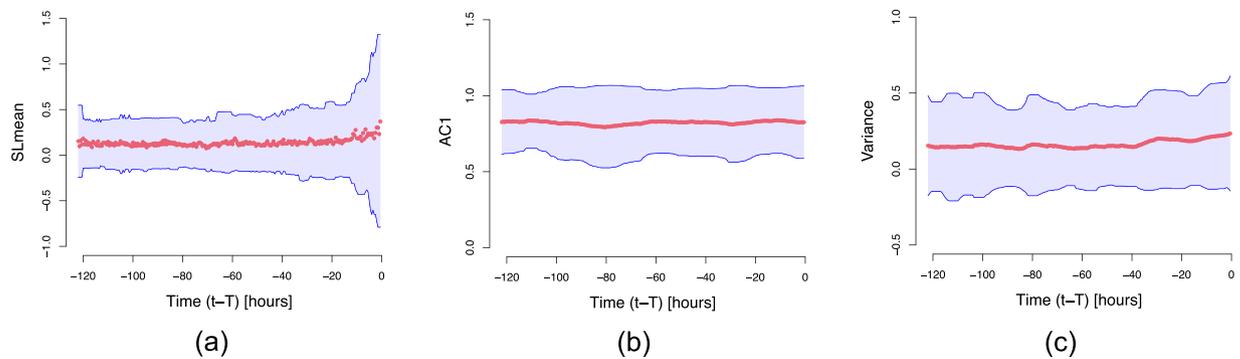

**Figure 5.** Componentwise mean (red dots) and ±standard deviation (blue lines) of (**a**) *SLMean* (**b**) *AC1* and (**c**) *VAR* for all septic shock patients prior to disease-onset (see 'SLMean as an early-warning indicator'); $T$ is the length of the time series (i.e. $\max(T^{1_c}, \ldots, T^{N_c})$, where $c=1$ represents septic shock condition and $N$ is the total number of septic shock patients), and $t-T$ is the time before the onset of septic shock. The number of samples per time point could be different due to the heterogeneous length of hospitalization (see 'Data source'). As the maximum length of hospitalization was 144 hours, with a moving time-window length of 18 hours and an average window of 3 hours, the minimum value of $t-T$ was $-123$ hours.

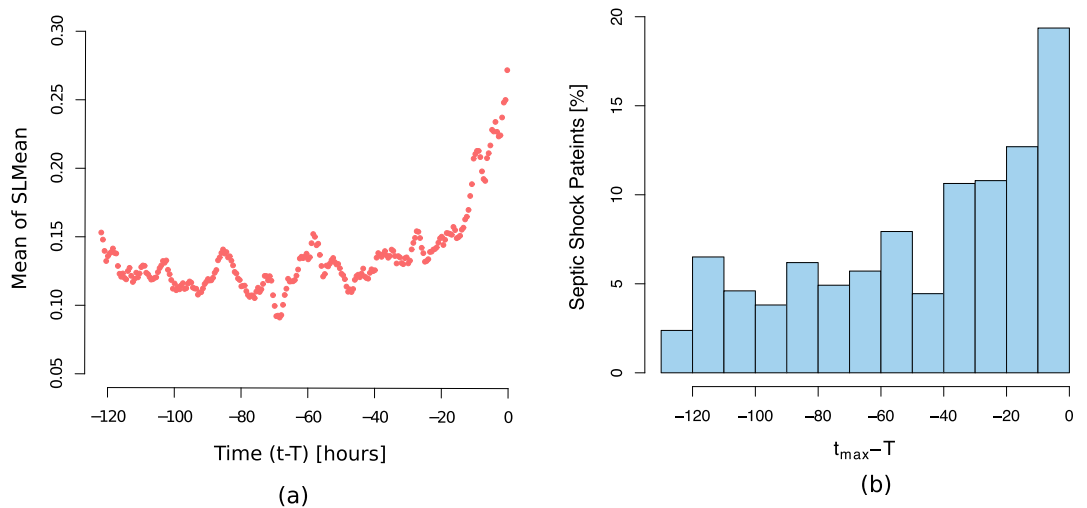

**Figure 6.** (**a**) Componentwise mean of *SLMean* for all septic shock patients prior to disease-onset (see 'SLMean as an early-warning indicator'); $T$ is the length of the time series, (**b**) Distribution of the times of critical transitions from the onset times of septicshock, i.e. $t_{max}^{1_c} - T^{1_c}, \ldots, t_{max}^{N_c} - T^{N_c}$, where $c=1$ represents septic shock condition and N is the total number of septic shock patients. $SLMean^{1_c}$ reaches a maximum at $t_{max}^{1_c}$.

**Robustness of the SSM model.** We assessed the robustness of our method to perturbations in the model parameters. We changed the length of the moving time window, $m \in (24, 30)$, and the number of trends in the SSM model ($e \in (4, 5)$) and compared the changes in the $t_{max}$ with respect to the reference setting, i.e., $m=36$ and $e=3$. The chosen values of $e$ are based on the assumptions described in 'State space model' ($e=3$ and $e \ll n$). The length of the moving time window was selected with regard to the average variables sampling rate (see Table 1, as well as the length of hospitalization in the ICU (see 'Data source'). The differences in $t_{max}$ due to the perturbations are summarized in online Supplementary Fig. S2. The zero median of such differences confirmed the robustness of our approach. Due to multiple similar high values *SLMean* in some patients, alteration of model parameters led to different $t_{max}$ values in these patients, which caused the outliers in online Supplementary Fig. S2.

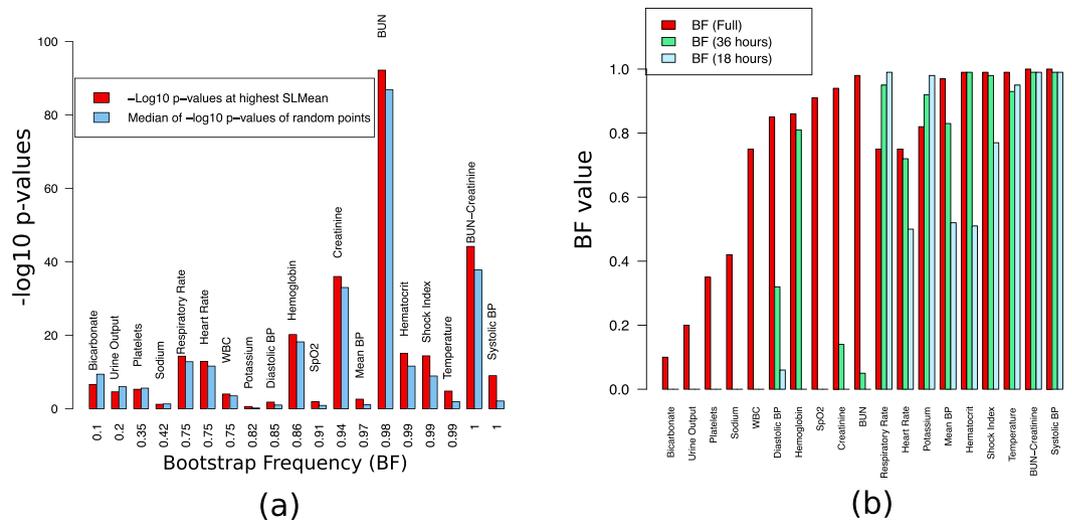

**Figure 7.** (**a**) A statistical significance test (see 'Data-sampling strategy with SLMean') was performed to test whether the values of the clinical variables at largest *SLMean* were able to differentiate septic shock patients from non-sepsis patients. The $-log_{10}(P\text{-}value)$ of each variable at $t_{max}$ was compared with the median $-log_{10}(P\text{-}value)$ of randomly selected points from the whole sequence. A bootstrap test (denoted as *BF Full*) was performed to quantify the number of times the p-values at $t_{max}$ are lower than those at $t_{random}$. (**b**) Another bootstrap test, performed by randomly sampling points from 18 hours and 36 hours windows prior to the onset of septic shock, tested whether low p-values at $t_{max}$ are an effect of time or a characteristic of regions with high *SLMean* values.

## Discussions

Healthcare can benefit from the analysis of continuously monitored health data, which are rapidly growing in quantity due to the increasing availability of long time series collected either by wearables or by monitoring systems such as those established in the ICU. However, significant challenges remain unresolved. A major drawback is the restriction of data availability to variables that are easy to collect by noninvasive sensors. These variables provide only correlated surrogates of the primary disease-driving processes. Hence, sensor signals are rarely specific on their own; advanced computational processing is typically necessary to identify relevant signals to improve therapy.

Focusing data analysis on the prediction and identification of critical transitions, i.e., instabilities in patient data, may complement established scoring methods in the classification of stable states. Although critical transitions differ qualitatively from scores in classifying stable states, the former method provides an independent assessment of health status. Because critical transitions are simply identified through the evolution of individual longitudinal time series, in contrast to established scores based on absolute variable values, markers for the detection of critical transitions are relatively robust to normalization and data standardization issues.

To identify such critical transitions in ICU patients, we applied the concept of surprise loss (SL), which was originally developed for determining instability in a model's forecasting ability in econometrics. We changed the model in the original SL approach to a multivariate SSM model to model two sources of variability, namely, the hidden underlying biological processes and the observables. Despite a multitude of interventions in the ICU, our moving average SL, *SLMean*, showed, on average, an increasing signal approximately 24 hours before the expert-annotated onset of septic shock (see Fig. 6a), thereby indicating its applicability as an early-warning indicator. We utilized such an indicator to devise a critical-transition-based data-sampling strategy for discriminating septic shock from non-sepsis patients. Additionally, through a bootstrap test (quantified through *BF(Full)*), the benefit of our method is shown with respect to a random data selection strategy (as summarized in Table 2 and Fig. 7a). Except for bicarbonate, urine output, platelets and sodium, the SL-based approach results in better p-values than the random strategy. In addition, we selectively sampled random data from 36 hours and 18 hours before the septic onset to compute *BF(36 h)* and *BF(18 h)*, respectively (see Fig. 7b and Supplementary Table S2). Such selective sampling evaluates whether merely sampling data close to the onset time of septic shock outperforms our method in distinguishing sepsis from non-sepsis. These new *BF* values seem to be well-preserved for most variables that have correspondingly high *BF(Full)*. Therefore, an SL-informed sampling strategy may improve the quality of patient classification and eventually enable the reduction of sample sizes.

Moreover, from a systems theory point of view, mechanisms that control the system in homeostasis begin to collapse around a critical transition or tipping point. Consequently, variables that are under tight control within stable states may be more sensitive to systemic variability around an unstable point. Our data analysis supports this hypothesis (see Fig. 7a): some variables under tight control, e.g., blood pressure and body temperature, showed significant improvement in p-values compared to random sampling. We compared our method with two other univariate early-warning measures for critical transitions in complex systems: temporal autocorrelation and variance[7,15,16,35]. As shown in Fig. 5, our method outperformed these estimators as an early-warning indicator for septic shock patients. Similarly, the p-values and *BF* of our method were also more favorable than those of the other methods (Table 2).

Conceptually, SL computation is based on the premise that the underlying system has a stable stationary state and that all observed deviations can be explained as responses to stochastic perturbations. The permissible amount of deviation is controlled by the system's robustness at the time of computation. As a result, SL-based analysis can forewarn of a "loss of stability" even before the underlying system has changed its state. In that sense, SL provides indicators similar to those from the analysis of critical slowing down[35]. One drawback is that local loss of robustness may neither result in a transition to another state nor indicate a new state. SL-based warning systems, in isolation, may thus lead to false alarms and could be improved by combining them with ML classifiers. Additionally, moving-window length restricts the capability of the SL-based warning system, and analysis can only be performed only when sufficient data have been collected. Hence, further evaluations must be performed towards utilization of SL-based analysis in practice. As a high SL is not specific and can be generated by any sudden event affecting the data, either errors in the monitoring system or health-related covariates, a robust characterization of the standard SL patterns characterizing control states is crucial. As sudden, high SL peaks can arise from sudden monitoring aberrations, we expect that a threshold-based alarm system might result in an unacceptable false positive rate. Therefore, emphasis should be placed on the characterization of SL patterns that are representative of the control state, eventually for each individual patient, followed by an AI-based pattern classifier. Effectively, this method will result in significant calibration times to setup the alarm system for each patient, such that effective training procedures for the learning of the control state patterns might be essential for transfer to clinical applications.

## Acknowledgements

The computing resources were granted by RWTH Aachen University under project rwth0260. S.S.S. was supported by funding from CompSE profile area, RWTH Aachen University. We wish to thank the anonymous reviewers whose constructive comments helped to improve the manuscript.

## Author Contributions

P.F.G. and S.S.S. developed the idea, conducted the research, and implemented the algorithms. J.S.B. helped in the preparation of the data and in proofreading of the article. R.D. and G.M. provided the clinical insights and interpreted the findings. A.S. supervised and supported the research project. All authors have reviewed the manuscript.

## Additional Information

**Competing Interests:** The authors declare no competing interests.

**Publisher's note:** Springer Nature remains neutral with regard to jurisdictional claims in published maps and institutional affiliations.